\documentclass[%
reprint,
 unsortedaddress,
 nofootinbib,
 amsmath,amssymb,
 aps,
 a4paper
]{revtex4-2}
\pdfoutput=1
\usepackage{url}
\usepackage{hyperref}

\usepackage{graphicx}
\usepackage{dcolumn}
\usepackage{bm}
\usepackage{slashed}
\usepackage{color}
\usepackage{physics}
\usepackage{comment}
\usepackage{esvect}
\usepackage[normalem]{ulem}
\usepackage{tikz}
\usetikzlibrary{quantikz2}

\begin{document}
\title{Directional search for light dark matter with quantum sensors}

\author{Hajime Fukuda}
\email{hfukuda@hep-th.phys.s.u-tokyo.ac.jp}
\affiliation{Department of Physics, The University of Tokyo, Tokyo 113-0033, Japan}

\author{Yuichiro Matsuzaki}
\email{ymatsuzaki872@g.chuo-u.ac.jp}
\affiliation{
Department of Electrical, Electronic, and Communication Engineering, Faculty of Science and Engineering, Chuo University, 1-13-27, Kasuga, Bunkyo-ku, Tokyo 112-8551, Japan
}
\author{Thanaporn Sichanugrist}
\email{thanaporn@hep-th.phys.s.u-tokyo.ac.jp}
\affiliation{Department of Physics, The University of Tokyo, Tokyo 113-0033, Japan}

\date{\today}

\begin{abstract}
The presence of dark matter (DM) stands as one of the most compelling indications of new physics in particle physics.
Typically, the detection of wave-like DM involves quantum sensors, such as qubits or cavities. The phase of the sensors is usually discarded as the value of the phase itself is not physically meaningful. However, the difference of the phase between the sensors contains the information of the velocity and direction of the DM wind. We propose a measurement protocol to extract this information from the sensors using quantum states. Our method does not require specific experimental setups and can be applied to any type of DM detector as long as the data from the detectors can be taken quantum mechanically. We also show that our method does not spoil the sensitivity of the DM detectors and is superior to the classical method based on the correlations of the DM signals between the detectors.
\end{abstract}

\maketitle
\newpage

\textit{Introduction.---}
Dark matter (DM) constitutes a significant fraction of the universe\,\cite{Planck:2018vyg}, yet its fundamental properties remain elusive. Among various candidates, ultra-light particles such as axions, axion-like particles, and dark photons are particularly well motivated~\cite{Arias:2012az}. As their masses are sufficiently small, these particles exhibit wave-like rather than particle-like behavior. 
Although their couplings to Standard Model (SM) particles are expected to be extremely weak, a wide range of experiments and observations are actively searching for signatures of such interactions.

DM comes to the Earth from all directions, but its flux is anisotropic. Due to the motion of the solar system through the galactic halo, an enhanced flux--often referred to as the ``DM wind''--is expected in the direction of the solar system's motion. This anisotropy is a distinctive feature of DM\,\cite{Spergel:1987kx}. For particle-like DM, numerous experimental efforts have focused on detecting this wind by measuring the recoil of target particles\,\cite{Spergel:1987kx,Lewin:1995rx,Morgan:2003qp,Morgan:2004ys,Morgan:2005sq,Green:2006cb,Bertone:2010zza,Miuchi:2023act}. By observing such recoils, one can determine the velocity distribution of the DM.

However, for wave-like DM, the situation is different, as the recoil of target particles is extremely small and generally undetectable. If the DM couples to SM particles via velocity-dependent interactions--such as the axion coupling to fermion spin\,(e.g. Ref.\,\cite{Garcon:2017ixh,Chigusa:2023roq,Chigusa:2023hmz})--the DM wind can be probed by changing the orientation of the apparatus. However, this approach is highly model-dependent, and the sensitivity is limited by the small DM velocity and unknown coupling strength, making independent measurement of the DM wind velocity challenging.
Alternatively, one could construct experimental apparatus with sizes comparable to the de Broglie wavelength of the DM\,\cite{Irastorza:2012jq,Knirck:2018knd}. For example, in cavity detectors, the axion-to-photon conversion probability depends on the direction of the axion wind. However, this requires specialized and often large setups, especially for lighter DM.

In this paper, we show that if the quantum states from DM detectors can be transferred over a distance and processed quantum mechanically, it is possible to independently and simultaneously measure both the velocity of the DM wind and the coupling between DM and SM particles. Our method is broadly applicable to any type of DM detector, regardless of the specific detection mechanism or the type of DM, if the detector state can be read out quantum mechanically.
Furthermore, we find that, in the absence of noise, this approach does not sacrifice the sensitivity of the detector; we extract additional information from the DM detectors.

The key idea of our proposal is using the quantum interference of the quantum sensors at distant positions. The information of the DM phase is generally encoded in the phase of the detector state. The value of the phase itself has little physical meaning, but the phase difference is sensitive to the momentum of the DM wind.
The \emph{classical} correlations of the DM signals between the detectors at distant positions have been discussed in Ref.\,\cite{Derevianko:2016vpm}, but we rather focus on the \emph{quantum} correlations of the DM signals; we measure the interference due to the DM wind by a non-local operator. Our method outperforms the classical correlations for weak signals.

Our proposal requires transferring quantum states between distant locations.
As we will discuss, the distance between the detectors is to be of the order of the de Broglie wavelength of the DM 
and we need to transfer the quantum states even over kilometers depending on the DM mass.
However, such operations are fundamental in quantum information science\,\cite{Bennett:1992tv} and used in various applications, such as the quantum key distribution\,\cite{Bennett:2014rmv}.
Indeed, quantum teleportation, which transfers quantum information between remote parties, has already been experimentally demonstrated by using optical photons\,\cite{PhysRevLett.80.1121}, solid-state qubits such as trapped atoms\,\cite{PhysRevLett.110.140403}, rare-earth-doped crystals\,\cite{Lago-Rivera:2022tzo}, superconducting qubits\,\cite{Qiu:2023gfq} and the nitrogen-vacancy (NV) center in diamond\,\cite{pfaff2014unconditional,hermans2022qubit}. For the distance, Ref.\,\cite{krutyanskiy2019light} demonstrates qubit-photon entanglement over 50 km.
Transfering quantum states may introduce additional channel noise, but this can be mitigated by entanglement distillation techniques\,\cite{campbell2008measurement}, which have also been experimentally demonstrated\,\cite{kalb2017entanglement}.
This capability is useful not only for quantum communication\,\cite{rozpkedek2019near} but also for quantum sensing\,\cite{degen2017quantum}.

\textit{Quantum mechanical signal from dark matter.---}
\label{sec:DM}
We first review how DM interacts with quantum detectors. For concreteness, we focus on superconducting qubits coupling directly to the DM field\,\cite{Chen:2022quj,Chen:2024aya}, but the discussion applies equally to other two-level systems such as NV centers\,\cite{Chigusa:2023roq} and trapped ions\,\cite{Ito:2023zhp}. Our approach is also applicable to other quantum sensors, including cavity detectors\,\cite{PhysRevLett.51.1415,ADMX:2019uok}, as long as the output is accessible quantum mechanically. We briefly comment on this at the end of the section.

Consider a qubit with ground and excited states $|0\rangle$ and $|1\rangle$. The Hamiltonian is:
\begin{align}
    H_0 = -\frac{\omega}{2} \sigma_z,
\end{align}
where $\omega$ is the energy splitting and $\sigma_i$ is the $i$-th Pauli matrix. The interaction with the DM field is modeled as
\begin{align}
    H_1 = 2 \eta \sigma_x \Phi(t), \label{eq:H1}
\end{align}
with the coupling $\eta$ and the DM field $\Phi(t)$. For a monochromatic DM field,
\begin{align}
    \Phi(t) = \Phi_0 \cos(Et + \varphi),
\end{align}
with $E = m(1 + v^2/2)$ the DM energy, $m$ the DM mass, $v$ 
its velocity, $\Phi_0$ 
the amplitude, and $\varphi$ 
the phase.

Assuming the detector is tuned ($\omega = E$), by the rotating wave approximation (RWA), the interaction Hamiltonian in the interaction picture becomes
\begin{align}
    H_I = \epsilon (\sigma_x \cos\varphi - \sigma_y \sin\varphi),
\end{align}
where $\epsilon \equiv \eta \Phi_0$. Starting from $|0\rangle$, after time $\tau$ the state evolves to
\begin{align}
    |\psi(\tau)\rangle \simeq |0\rangle - i \epsilon \tau e^{-i \varphi} |1\rangle, \label{eq:psi_pw}
\end{align}
for $|\epsilon \tau| \ll 1$. The probability to find the qubit in $|1\rangle$ is
\begin{align}
    p = |\langle 1 | \psi(\tau) \rangle|^2 \simeq (\epsilon \tau)^2, \label{eq:prob_pw}
\end{align}
which allows measurement of the DM interaction strength $\epsilon$.

Up to here, we have assumed that the DM field is oscillating with a single frequency. However, in reality, this may not be the case. In this paper, we assume the model adopted in Refs.\,\cite{Foster:2017hbq,Cheong:2024ose}, where the DM field is a sum of many oscillating fields with random phases:
\begin{align}
    \Phi(t) = \frac{\Phi_0}{\sqrt{N_{\rm DM}}} \sum_{i=1}^{N_{\rm DM}} \cos\qty[m\qty(1 + \frac{v_i^2}{2})t + \varphi_i],
    \label{eq:superposed_field}
\end{align}
where $N_{\rm DM}$ is the number of DM particles that make up the DM field, $v_i$ is the velocity of the $i$-th DM particle, and $\varphi_i$ is the phase of the $i$-th DM particle. 
Let the time we measure the DM field be $\tau$. 
For such $\Delta v$ that $m v_i \Delta v \tau \ll 1$, the oscillation frequencies of the DM field with velocity $v$ satisfying $|v - v_i| \lesssim \Delta v$ are almost the same, and we can add them up:
\begin{align}
    & \frac{\Phi_0}{\sqrt{N_{\rm DM}}}  \sum_{|v - v_i| \lesssim \Delta v} \cos\qty[m\qty(1 + \frac{v_i^2}{2})t + \varphi_i] \nonumber\\
    &= \Phi_0 \sum_{\text{direction}}\alpha_i \sqrt{f(\vec{v}_i) \Delta v^3} \cos\qty[m\qty(1 + \frac{v_i^2}{2})t + \Tilde{\varphi}_i], \label{eq:nearbysum}
\end{align}
where $\alpha_i$ is a random variable obeying the Rayleigh distribution, $P(\alpha_i) = \alpha_i e^{-\alpha_i^2/2}$, $f(\vec{v}_i)$ is the distribution function of the DM velocity, and $\Tilde{\varphi}_i$ is a random variable uniformly distributed in the range $[0, 2\pi)$.
The sum in the right-hand side is taken over the direction of the DM velocity.
Tuning the frequency of the detector to be equal to $m\qty(1 + \frac{\bar{v}^2}{2})$, where $\bar{v}$ is the average speed of the DM, we can again use the RWA and the interaction Hamiltonian is now
\begin{align}
    H_I &= \epsilon \sum_i \alpha_i \sqrt{\frac{f(\vec{v}_i)  \Delta v^3}{2}}     \nonumber \\
    &\times\left[ \sigma_x \cos\qty(\frac{m \Delta v_i^2}{2}t + \Tilde{\varphi}_i)  \right. \left.- \sigma_y \sin\qty(\frac{m \Delta v_i^2}{2}t + \Tilde{\varphi}_i) \right], \label{eq:HI_RWA_alpha}
\end{align}
where $\Delta v_i^2 \equiv v_i^2 - \bar{v}^2$.
Then, the state of the detector after $\tau$ is
\begin{align}
    |\psi(\tau)\rangle 
    &\simeq |0\rangle - i \epsilon \tau \sum_i \alpha_i \sqrt{\frac{f(\vec{v}_i) \Delta v^3}{2}}e^{-i\qty(\delta_i + \Tilde{\varphi}_i)} F\left(\delta_i\right) |1\rangle, \label{eq:specastate}
\end{align}
where $F(x) = \frac{\sin x}{x}$ and $\delta_i = \frac{m \Delta v_i^2}{4} \tau$.
The probability of the detector to be $|1\rangle$ is given by
\begin{align}
    p &= \left| \langle\psi(\tau) | 1\rangle \right|^2 \simeq \epsilon^2 \tau^2 \sum_{i} \alpha_i^2 \frac{f(\vec{v}_i) \Delta v^{3}}{2} F^2\left(\delta_j\right). \label{eq:prob_stat}
\end{align}
We have used the fact that the random phases $\Tilde{\varphi}_i$ are uncorrelated, and the cross terms with $i \neq j$ average to zero. Taking the average over the random amplitude $\alpha_i$, we obtain $\langle p \rangle \simeq \epsilon^2 \tau^2$, where we assume that $m \bar{v}^2 \tau \lesssim 1$ and $F(x) \simeq 1$, and we have used the fact that the distribution function $f(\vec{v})$ is normalized, $\int d^3v f(\vec{v}) = 1$. The average probability gives the same result as the case with a single frequency, Eq.\,\eqref{eq:prob_pw}.

We can take the average over the random phase $\varphi_i$
in the density matrix formalism. 
There, we treat the detector state as an ensemble of states, each having evolved under the interaction Hamiltonian with DM field, Eq.\,\eqref{eq:superposed_field}, with random phases; the final state of the detector is, instead of $|\psi(\tau)\rangle$, given by
\begin{align}
    \rho(\tau) \simeq \int \qty(\prod_i \frac{d\varphi_i}{2\pi}) U(\tau) \ket{0}\bra{0} U^\dagger(\tau), \label{eq:density_formalism}
\end{align}
where $U(\tau) \equiv \text{T} \exp\left[-i \int_0^\tau dt H_I(t)\right]$ depends on the random phases $\varphi_i$.
Again, we can take the partial sum over the DM velocities $v_i$, rewriting the integration over $\varphi_i$ in terms of $\alpha_i$ and $\tilde{\varphi}_i$. Performing the integration, we obtain
\begin{equation}
    \rho(\tau) \simeq |0\rangle\langle 0| + \epsilon^2 \tau^2 \int d^3v f(\vec{v}) F^2\left(\frac{m \qty(v^2 - \bar{v}^2)}{4} \tau\right)|1\rangle \langle 1|. 
\end{equation}
Averaging over the random phases removes all phase information when considering only the state of a single qubit. Consequently, no useful information from the phase of the DM can be extracted in this case. However, \emph{the difference} of the phases can be observable, as is shown in the next section.

Up to this point, we have assumed two-level qubits as the DM detectors. 
However, the information of the DM phase is also encoded in other types of DM detectors, such as resonant cavities\,\cite{Sikivie:1983ip}. Therefore, if we can take the data from the detectors quantum mechanically, we can apply our proposal to these detectors as well. As an example, we discuss the case of the cavity detectors in App.\,\hyperref[app:cavity]{A}.

\textit{Measurement protocol.---}
\label{sec:protocol}
Now, we present our measurement protocol to extract information about the DM wind using quantum interference between two spatially separated qubits. Let the qubits be positioned at $\vec{x}_1$ and $\vec{x}_2$, separated by $\vv{\Delta r} = \vec{x}_2 - \vec{x}_1$. To account for spatial dependence, we replace the random phase $\varphi$ in the DM field with $\varphi - \vec{k} \cdot \vec{x}$, where $\vec{k} = m \vec{v}$ is the DM wave vector. We assume $\Delta v$ is so small that $m \Delta v \Delta r \ll 1$ to ignore the phase difference.

Suppose we initialize both qubits in the ground state $\ket{00}$. After $\tau$, the state evolves to
\begin{align}
    \rho(\tau) \simeq \ketbra{00}{00} + \biggl[ \epsilon^2 \tau^2 \int d^3v\, f(\vec{v}) F^2\left(\frac{m (v^2 - \bar{v}^2)}{4} \tau\right)  \nonumber \\
    \times \left( |10\rangle + e^{i \vec{k} \cdot \vv{\Delta r}} |01\rangle \right) \left( \langle 10| + e^{-i \vec{k} \cdot \vv{\Delta r}} \langle 01| \right) \biggr].
    \label{eq:state_after_tau}
\end{align}
As before, the phases are averaged, but the difference of the phases, $\vec{k}\cdot\vv{\Delta r}$, remains in the density matrix.

We first perform a projective measurement to select events where exactly one qubit is excited without specifying which one. The corresponding projection operator is
\begin{align}
    P_1 = |10\rangle \langle 10| + |01\rangle \langle 01|. \label{eq:P1_operator}
\end{align}
The probability of this outcome is
\begin{align}
    p_1 = \mathrm{Tr}[P_1 \rho(\tau)] \simeq 2\epsilon^2 \tau^2 \int d^3v\, f(\vec{v}) F^2\left(\frac{m (v^2 - \bar{v}^2)}{4} \tau\right),
    \label{eq:prob_p1}
\end{align}
which is twice as large as that of the single-qubit result. The post-measurement state is
\begin{align}
    &\tilde{\rho}(\tau) = \frac{P_1 \rho(\tau) P_1}{\mathrm{Tr}[P_1 \rho(\tau)]} \nonumber \\ 
    &\simeq \frac{1}{2} \int d^3v\, f(\vec{v}) \left( |10\rangle + e^{i \vec{k} \cdot \vv{\Delta r}} |01\rangle \right) \left( \langle 10| + e^{-i \vec{k} \cdot \vv{\Delta r}} \langle 01| \right).
    \label{eq:density_after_P}
\end{align}
Here, we have neglected $F$ by assuming $m (v^2 - \bar{v}^2) \tau/4 \ll 1$ for relevant $v$.

To extract information about the DM wind, we measure the following operator:
\begin{align}
    M = -i |01\rangle \langle 10| + i |10\rangle \langle 01|. \label{eq:M_operator}
\end{align}
(See App.\,\hyperref[app:quantum_circuit]{B} for the quantum circuit implementation.)
The expectation value is
\begin{align}
    \overline{M} = \mathrm{Tr}[\tilde{\rho}(\tau) M] = -\int d^3v\, f(\vec{v}) \sin \left(m\vec{v} \cdot \vv{\Delta r}\right).
    \label{eq:Mbar}
\end{align}
Since operators $P_1$ and $M$ are applied to the two qubits at a distance, we need non-local gates to measure these observables. The state transfer over distance using quantum teleportation also helps perform such operators.

We adopt the standard halo model for the DM velocity distribution\,\cite{Baxter:2021pqo}:
\begin{align}
    f(\vec{v}) \propto \frac{1}{\pi^{3/2} v_0^3}  \exp & \left[-\frac{(\vec{v} + \vec{v}_\text{obs}(t))^2}{v_0^2}\right] \nonumber \\
    &\times \Theta(v_{\text{esc}} - |\vec{v} + \vec{v}_\text{obs}(t)|),
\end{align}
where $v_0$ is the local standard of rest velocity, $v_{\text{esc}}$ is the escape velocity, $\vec{v}_\text{obs}(t) = \vec{v}_0 + \vec{v}_\odot + \vec{v}_\oplus(t)$ includes the solar peculiar velocity $\vec{v}_\odot$ and Earth's orbital velocity $\vec{v}_\oplus(t)$, and $\Theta(x)$ is the Heaviside function.

\begin{figure*}[tb]
  \centering
  \raisebox{0.3mm}{\includegraphics[width=0.45\textwidth]{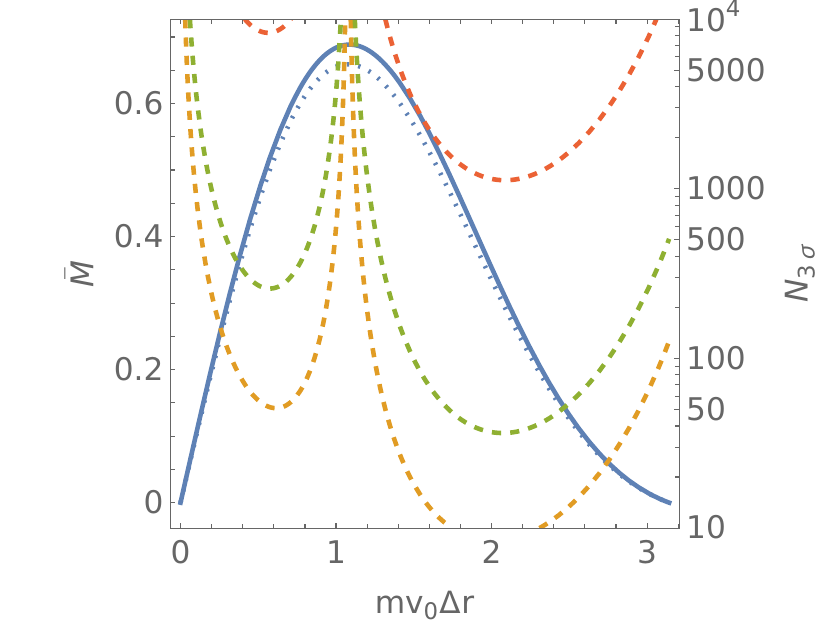}}
  \hspace{0.02\textwidth}
  \includegraphics[width=0.469\textwidth]{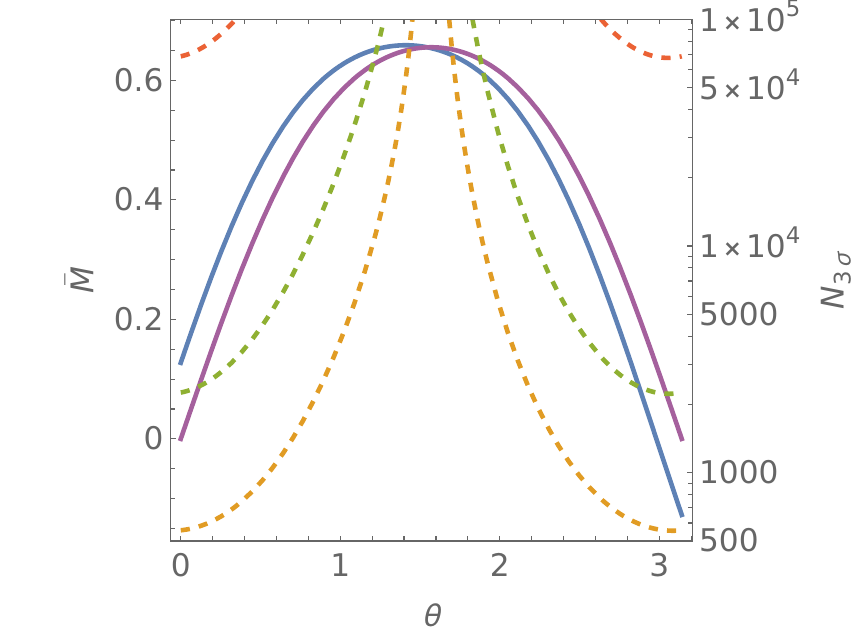}
  \caption{Left: The blue solid (dotted) line shows
  value of $\overline{M}$ as a function of $m v_0 \Delta r$ using numerical integration (analytic approximation) (left axis). The dashed lines show the number of measurements needed to achieve $v_{0}/\delta v_{0} = 3$ (right axis). The orange, green, and red lines correspond to the noise rate, $c=0, 2\epsilon^2\tau^2$, and $20 \epsilon^2\tau^2$, respectively. 
  Right: The solid lines show the value of $\overline{M}$ as a function of $\theta$, the angle between $\Delta r $ and the direction to the galactic center within the galactic plane (left axis). The blue (purple) line corresponds to the case with (without) the annual modulation of the DM wind. The dashed lines show the number of measurements needed to achieve $v_{\text{obs},x}/\delta v_{\text{obs},x} = 3$ (right axis). The color means the same as the left figure. (See Ref.~\cite{Fukuda_dataset2025} for the dataset.)}
  \label{fig:sensitivity}
\end{figure*}

Neglecting the escape velocity, the analytic expression for $\overline{M}$ is
\begin{align}
    \overline{M} = \exp\left(-\frac{m^2 v_0^2 \Delta r^2}{4}\right) \sin\left(m \vec{v}_\text{obs} \cdot \vv{\Delta r}\right).
\end{align}
The sensitivity is maximized when the separation $\Delta r$ is of the order of the DM de Broglie wavelength, $\lambda = 2\pi/(m v_0)$. The direction of $\vv{\Delta r}$ determines the velocity component being probed: parallel to $\vec{v}_0$ to access the local standard of rest velocity, or perpendicular to capture the annual modulation due to Earth's motion.

The sensitivity to the $i$-th component of $\vec{v}_\text{obs}$ is
\begin{align}
    \delta v_{\text{obs},i} = \frac{\sqrt{1 - \overline{M}^2}}{\sqrt{N}} \left| \frac{d \overline{M}}{d v_{\text{obs},i}} \right|^{-1},
\end{align}
where $N$ is 
the number of measurements of $M$ under the constraint that the post-selection with $P_1$ is done\,\cite{degen2017quantum}. This expression assumes $N$ is sufficiently large\,\cite{sugiyama2015precision}.

So far, we have neglected quantum noise. To include it, we consider depolarization noise acting on each qubit independently before the measurements. This models various noise types, including local noises such as environmental noise and infidelities in the state transfer. The state in Eq.\,\eqref{eq:state_after_tau} becomes
\begin{align}
    \rho'(\tau) \simeq \rho(\tau) + \frac{c}{2} \left( |10\rangle\langle 10| + |01\rangle\langle 01| \right),
\end{align}
where $c \ll 1$ is the depolarization rate. The probability for $P_1$ is now $p_1' \simeq p_1 + c$, and the post-selected state is
\begin{align}
    \tilde{\rho}'(\tau) = \frac{p_1}{p_1 + c} \tilde{\rho}(\tau) + \frac{c/2}{p_1 + c} \left( |10\rangle\langle 10| + |01\rangle\langle 01| \right).
\end{align}
The expectation value of $M$ is suppressed:
\begin{align}
    \overline{M}' \simeq \frac{2\epsilon^2 \tau^2}{c + 2\epsilon^2 \tau^2} \, \overline{M}.
\end{align}
Other noise types similarly reduce sensitivity.

Fig.~\ref{fig:sensitivity} illustrates the sensitivity of our protocol to the DM wind. Here, we follow the conventions of Ref.\,\cite{Baxter:2021pqo}, expressing the quantities in the galactic rectangular coordinates. 
In the left panel, we align $\vv{\Delta r}$ with the direction of $\vec{v}_0$ (the $y$-axis) and vary $\Delta r$. We ignore $v_\odot$ and $v_\oplus(t)$ for simplicity here.
The analytic (solid) and numerical (dotted) results for $\overline{M}$ agree well. We therefore use the analytic solution ignoring the escape velocity for the other results.
Dashed lines indicate the number of measurements $N_{3\sigma}$ required for $v_{0}/\delta v_{0}=3$. Optimizing $\Delta r$ allows for efficient DM wind detection with a manageable number of measurements.

In the right panel, we set $m\Delta r = 1/v_0$ and vary the angle $\theta$ between $\vv{\Delta r}$ and the direction to the galactic center (the $x$-axis) within the galactic plane (the $xy$-plane), choosing $t$ to maximize the $x$-component of $\vec{v}_\text{obs}(t)$. The blue (purple) line shows $\overline{M}$ with $\vec{v}_\text{obs} = \vec{v}_0 + \vec{v}_\odot + \vec{v}_\oplus$ ($\vec{v}_\text{obs} = \vec{v}_0$). Dashed lines show $N_{3\sigma}$ for $v_{\text{obs},x}/\delta v_{\text{obs},x} = 3$ for the blue line, demonstrating sensitivity to the DM wind direction.

Both panels include the effect of depolarization noise: as the noise rate $c$ increases, more measurements are needed, but the sensitivity to the DM wind is retained even for $c \gg \epsilon^2 \tau^2$. While these results assume a fixed DM wind direction, in practice, Earth's rotation induces diurnal modulation\,\cite{Bertone:2010zza}, which can be resolved with sufficient measurement statistics.

\textit{Comparison with classical correlations.---}
\label{sec:comparison}
In the previous section, we demonstrated the DM wind can be measured via quantum interference between qubits with non-local operations. An alternative approach is using classical correlations with local measurements\,\cite{Derevianko:2016vpm}. Here, we compare the two methods and show that our quantum protocol with non-local operations
offers superior sensitivity for weak signals.

The local method using classical correlations
considers detectors measuring continuous DM field values. For weak signals, the detector can be modeled as a single two-level system, with field operators $\sigma_x$ and $\sigma_y$. For a monochromatic DM field, $\langle \sigma_x \rangle$ and $\langle \sigma_y \rangle$ encode the DM amplitude and phase. The DM wind information can be extracted from the two-point correlation:
\begin{align}
I \equiv \mathrm{Tr} [\sigma^{(1)}_y \sigma^{(2)}_x  \rho(\tau)] \simeq 2\epsilon^2\tau^2 \int d^3v\, f(\vec{v})\sin(m \vec{v}\cdot \vv{\Delta r}),
\end{align}
where the superscripts at Pauli matrices denote the index of the detectors.
The two-point correlation is of the same order as the quantum expectation value $\overline{M}$ in Eq.\,\eqref{eq:Mbar}, taking into account that we postselect states to measure $\overline{M}$.

The key difference lies in the measurement uncertainty. For $N_I$ repetitions, the uncertainty in $I$ is
\begin{equation}
\delta I  = \frac{\sqrt{\text{Tr}\qty[\qty(\sigma^{(1)}_y \sigma^{(2)}_x)^2 \rho(\tau)] - I^2}}{\sqrt{N_I}} \simeq \frac{1}{\sqrt{N_I}} + \mathcal{O}(\epsilon^4\tau^4).
\end{equation}
The corresponding DM wind resolution is
\begin{equation}
    \delta v^{(I)}_{\text{obs},i} \sim \frac{\sqrt{N}}{\epsilon^2\tau^2\sqrt{N_I}} \left(\frac{1}{\sqrt{N}}\left|\frac{d \overline{M}}{dv_{\text{obs},i}}\right|^{-1}\right) .
\end{equation}
To match the sensitivity of the quantum protocol with non-local operations, the local method with classical correlation requires $N_I \sim (\epsilon \tau)^{-4} N$ measurements, where $N$ is the number of postselected quantum measurements. Since the postselection probability is $p_1 \simeq 2\epsilon^2\tau^2$, the total number of quantum measurements is $N^{(\rm total)} = N / p_1$. Thus, the ratio of required measurements is
\begin{equation}
    \frac{N_I}{N^{(\rm total)}} \sim \frac{1}{(\epsilon\tau)^2}.
\end{equation}
Therefore, in the weak-field limit ($|\epsilon \tau| \ll 1$), our quantum protocol requires far fewer measurements than the classical correlation method for the same sensitivity.

This scaling persists even without postselection. Measuring $M$ directly, we have the expectation value $\langle M \rangle = p_1 \overline{M}$ and uncertainty $\delta M \simeq \sqrt{p_1}$. Then,
\begin{align}
    \label{eq:delta_v_avg_M_direct}
    \delta v^{(M)}_{\text{obs},i} \simeq \frac{1}{\sqrt{(\epsilon \tau)^2 N_M}} \left|\frac{d \overline{M}}{dv_{\text{obs},i}}\right|^{-1},
\end{align}
where $N_M$ is the number of measurements. Again, the method to use classical correlation requires $1/(\epsilon^2 \tau^2)$ times more measurements. The essential advantage of our approach stems from using the non-local operator $M$; $M$ directly accesses the phase difference induced by the DM wind, which is non-local as well. Indeed, as is shown in App.\,\hyperref[app:qcrb]{C}, the scaling of the uncertainty, Eq.\,\eqref{eq:delta_v_avg_M_direct}, in terms of $N_M$ and $\epsilon \tau$, is optimal, as it saturates the quantum Cram\'er-Rao bound\,\cite{Paris:2008zgg}.

\textit{Conclusion and discussion.---}
\label{sec:conclusion}
In this work, we have introduced a quantum protocol to measure the DM wind using interference between spatially separated quantum sensors. Our method enables simultaneous extraction of both the DM interaction strength and the DM wind velocity with each detection event. By optimizing the sensor separation, the protocol can probe both the solar system's motion relative to the DM halo and the Earth's orbital velocity. We have shown that the approach remains robust even in the presence of significant noise, and while we focused on the standard halo model with the distribution, Eq.\,\eqref{eq:superposed_field}, our method is adaptable to more complex DM distributions. It is interesting to ask to what extent the DM distribution can be reconstructed from the measurement results. We leave this question to future work.
The scanning with various separations of qubit detectors could also help gain information about the DM distribution.

Although we assumed identical qubits for simplicity, the protocol is general: if each sensor has a different interaction strength with the DM field, the measurement outcome is simply rescaled, and the method remains valid. Even with systematic phase shifts or different detector types, as long as the phase offset is stable, the DM wind information can still be extracted.

Extending the protocol to more than two distant sensors, which is considered a quantum sensing network~\cite{komar2014quantum,eldredge2018optimal,proctor2018multiparameter,ge2018distributed,kasai2022anonymous,kasai2024direct}, offers further possibilities. For example, arranging sensors in an array with uniform phase differences allows for efficient phase estimation via quantum Fourier transform (QFT) techniques~\cite{Coppersmith:2002skh}. Recent work suggests that entanglement and quantum computing can enhance sensitivity to unknown-frequency signals such as DM~\cite{Allen:2025hdx}. The use of techniques such as the QFT could open up another way to use quantum computers for extracting additional information.
More sophisticated protocols, including those using entangled states like GHZ states~\cite{Greenberger1989,Giovannetti_2011,matsuzaki2011magnetic,chin2012quantum}, may further improve performance, though they may well require more sophisticated measurement strategies, such as Ref.\,\cite{PhysRevResearch.2.023052}, since the response of the GHZ state to the DM is different\,\cite{Chen:2023swh}. We leave the exploration of these directions to future work.


\hfill

\textit{Acknowledgments.---}
The work of TS was supported by the JSPS fellowship Grant No.\ 23KJ0678.
This work was supported by JSPS KAKENHI Grant Nos.\ 24K17042 [HF], and 25H00638 [HF].
This work was supported by
JST Moonshot (Grant Number JPMJMS226C), CREST
(JPMJCR23I5), JST, and Presto
JST Grant Number JPMJPR245B.
In this research work, HF used the UTokyo Azure (\url{https://utelecon.adm.u-tokyo.ac.jp/en/research_computing/utokyo_azure/}).

\hfill

\textit{Appendix A: Cavity detector response to the DM field.---}
\phantomsection
\label{app:cavity}
In the main text, we discuss the details of the response of the qubit detector to the DM field. In this appendix, as an example, we show that a similar response is obtained from the cavity detector as well. For simplicity, here, we consider monochromatic DM, while the extension to the superposition of plane waves can be performed similarly as in the main text.

The Hamiltonian of an electromagnetic field inside a cavity in the second quantization picture is
\begin{equation}
H_0= \omega_{\rm c} a^\dagger a
\end{equation}
with the interaction between the DM field and the cavity field given by
\begin{equation}
H_1=2 \epsilon (a+a^\dagger) \cos(E t+\varphi ), \label{eq:cavityH1}
\end{equation}
where $E=m\left(1+\frac{1}{2}v^2\right)$ with $m,v$ are the mass and velocity of DM, respectively, and $\varphi$ is the phase of the DM field.
The operators $a,a^\dagger$ are the annihilation and creation operators of cavity photons satisfying
\begin{equation}
    [a,a^\dagger]=1, \quad a \ket{0}_c=0,
\end{equation}
where $\ket{0}_c$ is the vacuum state of the cavity. 
We denote the interaction strength between this cavity mode and DM as $\epsilon$, which is proportional to the amplitude of the DM field. We note that the form of the interaction given by Eq.~\eqref{eq:cavityH1} is obtained for the cavity haloscope experiments (see, e.g., Ref.~\cite{Chen:2024aya}). When the interaction strength $\epsilon$ is small and we focus on the Hilbert space of the two lowest-energy states $\{\ket{0}_c,a^\dagger \ket{0}_c\}$, one can see that the system is equivalent to that of the qubit detector system interacting with DM. Therefore, the same response should be expected. 

Let us consider the situation where $\omega_c=E$, the interaction Hamiltonian is given by
\begin{equation}
    H_I=\epsilon (a e^{i\varphi} +a^\dagger e^{-i\varphi}),
\end{equation}
where we have used the RWA.
Then, assuming that the initial state of the cavity is $\ket{0}_c$, we obtain the detector state at time $\tau$ as
\begin{align}
    \ket{\psi(\tau)}&= \text{T} \exp[-i\int_0^\tau dt H_I] \ket{0}_c\\
    &\simeq \ket{0}_c-i\epsilon \tau e^{-i\varphi} a^\dagger\ket{0}_c \label{eq:finalcavity}
\end{align}
where we assume a small interaction strength, i.e., $|\epsilon \tau|\ll 1$. The response is the same as that obtained with the qubit detector, Eq.~\eqref{eq:psi_pw}, as expected. 

We also briefly comment on transferring the information of the cavity state to a qubit degree of freedom for the subsequent state transfer over distance or subsequent measurements. Let us consider the Jaynes--Cummings Hamiltonian~\cite{1443594} as the interaction between the cavity and the qubit:
\begin{equation}
    \Delta H=\lambda (\ket{0}\bra{1} e^{-i\omega_{\rm q} t}+\ket{1}\bra{0} e^{i\omega_{\rm q} t}) (a e^{-i\omega_{\rm c} t}+a^\dagger e^{i\omega_{\rm c} t}),
\end{equation}
where $\lambda$ is a constant and $\omega_{\rm q}$ is the (effective) qubit frequency. The frequency $\omega_{\rm q}$ can be tuned, e.g. for superconducting qubits, by applying an oscillating field interacting with the qubit and inducing an AC Stark shift~\cite{Autler:1955zz, PhysRevLett.94.123602}. For those equipped with a superconducting quantum interference device (SQUID), the frequency can be tuned by adjusting the external magnetic flux through the SQUID loop~\cite{PRXQuantum.2.040202}. When $\omega_{\rm q}$ is set to $\omega_{\rm c}$, using RWA, we obtain the interaction as
\begin{equation}
    \Delta H\simeq \lambda \left( \ket{0} \bra{1 } a^\dagger+\ket{1} \bra{0 } a \right)
\end{equation}
which leads to the transition;
\begin{equation}
    a^\dagger \ket{0}_c \otimes \ket{0} \rightarrow \ket{0}_c \otimes \ket{1}.
\end{equation}
One might apply it to transfer the cavity state, Eq.~\eqref{eq:finalcavity}, to the qubit state as
\begin{equation}
    \qty(\ket{0}_c-i\epsilon\tau e^{-i\varphi} a^\dagger\ket{0}_c) \otimes \ket{0} \rightarrow \ket{0}_c\otimes \qty(\ket{0}-i\epsilon \tau e^{-i\varphi} \ket{1}).
\end{equation}

\hfill

\begin{figure}[thb]
  \centering
  \begin{quantikz}
        \lstick{qubit 1}          & \ctrl{2} &          & \targ{} & &&\\
        \lstick{qubit 2}          &          & \targ{}  & & && \\
      \lstick{ancilla, $\ket{0}$} &  \targ{} & \ctrl{-1} & \ctrl{-2} & \gate{S^\dagger} & \gate{H} & \meter{}
  \end{quantikz}
  \caption{Quantum circuit for $M$ measurement.}
  \label{fig:m_circuit}
\end{figure}
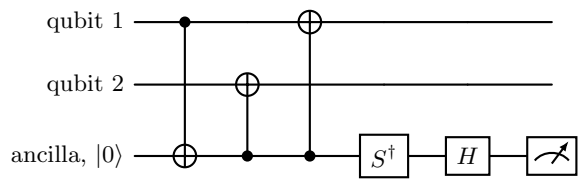

\textit{Appendix B: Quantum circuits for the measurements.---}
\phantomsection
\label{app:quantum_circuit}
In this appendix, we present the quantum circuits to perform the measurements $P_1$ and $M$ described in the main text. 
The circuit to perform the measurement $P_1$ (Eq.~\eqref{eq:P1_operator}) and $M$ (Eq.~\eqref{eq:M_operator}) are given by Figs.~\ref{fig:p_circuit} and \ref{fig:m_circuit}, respectively. For the convention of the quantum circuits and gates, we follow Ref.\,\cite{Nielsen:2012yss}.

\begin{figure}[thb]
  \centering
  \begin{quantikz}
        \lstick{qubit 1}        & \ctrl{2} &      &\\
        \lstick{qubit 2} &          & \ctrl{1} &  \\
      \lstick{ancilla, $\ket{0}$}     &  \targ{} & \targ{}  & \meter{}
  \end{quantikz}
  \caption{Quantum circuit for $P_1$ measurement.}
  \label{fig:p_circuit}
\end{figure}
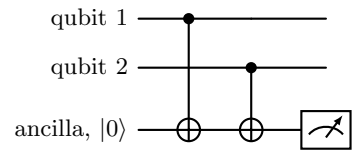

In the following, we derive the state of the qubits after passing through each circuit, demonstrating that the circuits indeed perform the desired operations.
The general state of 2 qubits with an ancilla qubit in the ground state can be written as
\begin{equation}
\ket{\psi}=\left(a\ket{00}
+b\ket{01}
+c\ket{10}
+d\ket{11} \right) \otimes \ket{0}_{\rm an},
\end{equation}
with $a,b,c,d$ as constants and we have added the subscript to the rightmost state to clarify it is for the ancilla qubit.
First, consider passing it through the circuit given by Fig.~\ref{fig:p_circuit}. There, we obtain the state before the measurement of the ancilla qubit as
\begin{equation}
    \ket{\psi_{P_1}}=\left( a\ket{00}+d\ket{11} \right)\otimes \ket{0}_{\rm an}+\left( b\ket{01}+c\ket{10} \right) \otimes \ket{1}_{\rm an},
\end{equation}
giving the desired projection operator $P_1$ when selecting only the $\ket{1}_{\rm an}$ outcome of the measurement of the ancilla qubit.

Next, let us consider the measurement $M$ given by Fig.\,\ref{fig:m_circuit}. This circuit works as $M$ measurement only when the state is projected by $P_1$. Thus, let us consider the general state with $a=d=0$. The final state before the measurement of the ancilla qubit is
\begin{align}
    \ket{\psi_M}= \frac{1}{\sqrt{2}} (b-ic)\ket{01} \otimes \ket{0}_{\rm an}  +\frac{1}{\sqrt{2}} (b+ic)\ket{01} \otimes \ket{1}_{\rm an}.
\end{align}
We may assign $M=+1$ for the $\ket{0}_{\rm an}$ outcome and $M=-1$ for the $\ket{1}_{\rm an}$ outcome, respectively.

\hfill

\textit{Appendix C: Quantum and Classical Fisher information for the DM wind measurement.---}
\label{app:qcrb}
In this appendix, we derive the quantum Fisher information (QFI) for the DM wind measurement, which gives the quantum Cram\'er-Rao bound (QCRB) for the uncertainty of the DM wind velocity. We show that Eq.\,\eqref{eq:delta_v_avg_M_direct} saturates the QCRB in terms of the number of measurements and the interaction strength. We also calculate the classical Fisher information for the separate measurement of each qubit and show that it is smaller than the QFI by a factor of $\epsilon^2 \tau^2$.

First, let us ignore the escape velocity for the DM velocity distribution and ignore $F(x)$ in Eq.\,\eqref{eq:state_after_tau} for simplicity. The density matrix can be written as
\begin{align}
    \rho(\tau) =& \ketbra{00}{00} + \epsilon^2 \tau^2 \Bigl[\ketbra{10}{10} + \ketbra{01}{01}  \nonumber \\
    &  + e^{-\Delta_0}\qty(e^{-i \theta} \ketbra{01}{10} + e^{i \theta} \ketbra{10}{01})\Bigr],
\end{align}
where $\Delta_0 = \frac{1}{4} m^2 v_0^2 \Delta r^2$ and $\theta = m \vec{v}_\text{obs} \cdot \vv{\Delta r}$. 
This state can be rewritten as
\begin{align}
    \rho(\tau) &= \ketbra{00}{00} + p_+ \ketbra{\psi_+}{\psi_+} + p_- \ketbra{\psi_-}{\psi_-},
\end{align}
where
\begin{align}
    p_\pm &= \epsilon^2 \tau^2(1 \pm e^{-\Delta_0}), \\
    \ket{\psi_\pm} &= \frac{1}{\sqrt{2}} \qty(\ket{10} \pm e^{-i \theta} \ket{01}).
\end{align}
Using this decomposition, we can calculate the QFI for the parameter $v_{\text{obs},i}$ as
\begin{align}
    \mathcal{F}_Q &= 2 \sum_{i,j=\pm} \frac{\qty|\bra{\psi_i} \partial_{v_{\text{obs},i}} \rho(\tau) \ket{\psi_j}|^2}{p_i + p_j} \sim \epsilon^2 \tau^2.
\end{align}

For $N_M$ measurements, the uncertainty of $v_{\text{obs},i}$ is given by the QCRB as
\begin{align}
    \delta v_{\text{obs},i} \geq \frac{1}{\sqrt{\mathcal{F}_Q N_M}} \sim \frac{1}{\epsilon\tau \sqrt{N_M}}.
\end{align}
Indeed, Eq.\,\eqref{eq:delta_v_avg_M_direct} gives the same scaling in terms of $N_M$ and $\epsilon \tau$.

Next, let us consider the classical Fisher information (CFI) for the separate measurement of each qubit. Using the detector state, Eq.\,\eqref{eq:specastate}, the probability of obtaining the outcome $s_a = 0,1$ when measuring $\sigma_a$ ($a=x,y$) is
\begin{align}
    P(s_x) &= \frac{1}{2} - (-1)^{s_x} \epsilon \tau \sum_i \alpha_i \sqrt{\frac{f(\vec{v}_i )\Delta v^3}{2}} \sin(\delta_i + \tilde{\varphi}_i), \\
    P(s_y) &= \frac{1}{2} - (-1)^{s_y} \epsilon \tau \sum_i \alpha_i \sqrt{\frac{f(\vec{v}_i )\Delta v^3}{2}} \cos(\delta_i + \tilde{\varphi}_i),
\end{align}
where we drop $F(\delta_i)$ by assuming $|\delta_i| \ll 1$ for relevant $v_i$. The probability of obtaining the outcome $(s_x^{(1)}, s_y^{(2)})$ when measuring $\sigma_x^{(1)}$ and $\sigma_y^{(2)}$ is
\begin{align}
    P(s_x^{(1)}, s_y^{(2)}) &= \sum_{\alpha_i} \int \qty(\prod_{\phi_i} \frac{d \phi_i}{2\pi}) P(s_x^{(1)}) P(s_y^{(2)}) \nonumber \\
    &\simeq \frac{1}{4} - (-1)^{s_x^{(1)} + s_y^{(2)}} \frac{\epsilon^2 \tau^2}{2} \overline{M},
\end{align}
where we retain terms up to $\epsilon^2 \tau^2$.
The classical Fisher information for $v_{\text{obs},i}$ is
\begin{align}
    \mathcal{F}_C &= \sum_{s_x^{(1)}, s_y^{(2)}} \frac{\qty|\partial_{v_{\text{obs},i}} P(s_x^{(1)}, s_y^{(2)})|^2}{P(s_x^{(1)}, s_y^{(2)})} \sim \epsilon^4 \tau^4.
\end{align}
Indeed, the information obtained by the separate measurement of each qubit is smaller than the QFI, which is the information obtained by our protocol, by a factor of $\epsilon^2 \tau^2$.

\bibliography{papers}


\end{document}